\documentclass[aps,prapplied,reprint,amsmath,amssymb,
               floatfix,nofootinbib,superscriptaddress]{revtex4-2}

\usepackage{graphicx}
\usepackage{bm}
\usepackage{algpseudocode}
\usepackage{tikz}
\usetikzlibrary{arrows.meta,positioning,calc,fit}
\usepackage[colorlinks=true,allcolors=blue]{hyperref}

\definecolor{RevTwo}{rgb}{0.00,0.00,0.00}
\definecolor{RevThree}{rgb}{0.00,0.00,0.00}
\definecolor{RevFour}{rgb}{0.00,0.00,0.00}
\definecolor{RevAll}{rgb}{0.00,0.00,0.00}
\definecolor{RevRoundTwo}{rgb}{0.60,0.00,0.60}
\definecolor{FixRed}{rgb}{0.85,0.10,0.10}

\usepackage{algpseudocode}

\newcounter{algorithm}
\renewcommand{\thealgorithm}{\arabic{algorithm}}
\newenvironment{alg}[1][t]
  {\begin{figure}[#1]\refstepcounter{algorithm}%
   \hrule height0.8pt\vspace{3pt}\footnotesize}
  {\vspace{2pt}\hrule height0.8pt\end{figure}}
\newcommand{\algcaption}[1]{%
  {\footnotesize\textbf{Algorithm~\thealgorithm.}\ #1\par}%
  \vspace{3pt}\hrule height0.4pt\vspace{3pt}}

\algrenewcommand\algorithmicindent{0.9em}
\algrenewcommand{\alglinenumber}[1]{\footnotesize#1:}
\algrenewcommand\algorithmiccomment[1]{\hfill\(\triangleright\) \textit{#1}}

\newcommand{\CFL}{S}
\newcommand{\Ex}{{E_x}}  \newcommand{\Ey}{{E_y}}  \newcommand{\Ez}{{E_z}}
\newcommand{\Hx}{{H_x}}  \newcommand{\Hy}{{H_y}}  \newcommand{\Hz}{{H_z}}
\newcommand{\dpl}{\delta^{+}}
\newcommand{\dmi}{\delta^{-}}
\newcommand{\Nz}{N(z)}

\begin{document}

\title{Ternary-Valued FDTD Method for Nonlinear Optics:\\
       Third-Harmonic Generation on a Multiplier-Free Solver}

\author{Ivan S.~Maksymov}
\email{ivan.maksymov@gmail.com}
\affiliation{Seymour Research Laboratories, Seymour, VIC 3660, Australia}

\date{\today}

\begin{abstract}
I show that a finite-difference time-domain (FDTD) solver with field variables restricted to the ternary alphabet $\{-1,0,+1\}$ can reproduce nonlinear optical physics quantitatively on both CPUs and field-programmable gate arrays (FPGAs). Building on the general ternary-valued FDTD algorithm, in which each Yee cell stores an integer state that accumulates the discrete curl of neighbouring symbols without being drained, I extend the scheme to inhomogeneous and Kerr media and to Mur absorbing boundaries, without performing any multiplication operations. Dielectric contrast is implemented as a multiplexer, since the difference between ternary symbols takes only a few values, the instantaneous Kerr nonlinearity as a coefficient look-up table and the Mur boundary coefficients as canonical signed-digit shift--add chains. The scheme reproduces the third-harmonic conversion efficiency of a floating-point reference to within $1.4\%$. A symmetry of the governing physics, expressed through the even harmonics forbidden by centrosymmetric $\chi^{(3)}$, provides a reference-free self-check for both ternary and floating-point models. The complete solver fits on a small, low-cost FPGA and contains no multiplier anywhere in its datapath, placing nonlinear optical modelling within reach of embedded and autonomous platforms rather than of high-performance computers.
\end{abstract}

\maketitle

% ====================================================================
\section{Introduction}
% ====================================================================

The growth of large-scale machine learning systems has increased rather than eliminated the need for first-principles electromagnetic solvers. Surrogate models---physics-informed neural networks~\cite{Rai19, Qi23} and neural operators among them~\cite{Ran21, Nam26}---are for the most part trained on data that a conventional solver must first produce, and even those trained from the governing equations alone are validated against data produced by a conventional solver. They interpolate reliably within the distribution of geometries, materials and excitations they
have seen, and degrade in ways that are difficult to bound outside it.
\begin{figure*}[t]
\centering
\includegraphics[width=\textwidth]{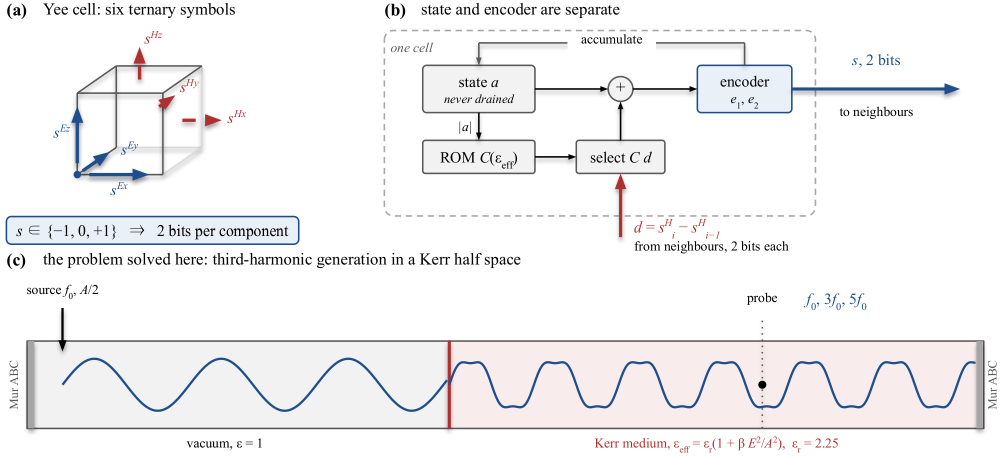}
\caption{\label{fig:concept}%
  Ternary-valued FDTD, from the 3D scheme to the problem
  solved here.
  (a)~A Yee cell. All six field components reach their neighbours as single ternary symbols $s\in\{-1,0,+1\}$, two bits rather than a word:~electric on the edges (solid), magnetic on the faces (dashed), as on a conventional staggered grid.
  (b)~One cell's electric update, in the 1D reduction. The
  state is never drained:~the encoder reads it without subtracting the quantiser threshold, so the full-precision field stays inside the cell; only two-bit symbols cross the boundary (dashed). Because the symbol difference $d$ takes five values here, Eq.~\eqref{eq:fiveway}, the product $C\,d$ is a multiplexer and $C$, including its Kerr dependence, a table lookup; no multiplier appears in the design. ROM:~read-only memory, a lookup table that is read but never written at run time.
  (c)~The test problem:~a vacuum half space abutting a Kerr medium, $\varepsilon_{r}=2.25$, driven at $f_{0}$ from the left, terminated by first-order Mur boundaries and probed inside the dielectric. The waveform is drawn rather than simulated, as $\sin\theta+0.17\sin3\theta$ with the wavelength shortened by $n=\sqrt{\varepsilon_{r}}=1.5$. Two terms suffice since $\chi^{(3)}$ generates only odd harmonics and the pump is undepleted. The ratio $0.17$ is the measured $a_{3}/a_{1}$ (Table~\ref{tab:harm}).}
\end{figure*}

In conventional electromagnetic solvers, such as finite-difference time-domain (FDTD) approaches applied to Maxwell's equations~\cite{Taf80, Yee66, Taflove, Johnson}, by contrast, the error is controlled by the discretisation order, providing the same guarantee for a structure never simulated before as for one in the training set. This is precisely what device qualification and certification require~\cite{IEEE}. Moreover, machine-learning workflows are voracious consumers of forward solutions:~inverse design and topology optimisation require a full Maxwell solution at every iteration~\cite{Mol18, Jen11, Min20}, and the generation of a training corpus requires one per sample, so that a single study routinely consumes $10^{3}$--$10^{6}$ independent solutions. The practical question is therefore not whether first-principles solvers remain necessary, but how cheaply each forward solve can be performed. This is the question addressed in this work.

FDTD solvers and adjacent techniques~\cite{Taf80, Yee66, Taflove, Johnson} are usually limited by memory bandwidth rather than by arithmetic, and the quantity moved between cells is a wide floating-point word. This has motivated a long line of reduced-precision and hardware-accelerated approaches~\cite{yu2013valu, Pfeifle2015, Mansoori, Yao26, Mak_FDTD}. The extreme case is a field variable of two bits~\cite{Mak26_2D}, which reduces inter-cell traffic by more than an order of magnitude and removes the multiplier from the datapath entirely. The same pressure has produced ternary weights in machine learning, where a parameter takes only the values $-1$, $0$ and $+1$, enabling a two-bit representation~\cite{Ma24, Par25, Qia26,dade2026}.

A particular difficulty in carrying this over to FDTD is that na\"ive quantisation of a Yee update destroys the scheme:~if the quantiser threshold is subtracted back out of the accumulator, the time integration is removed with it and what remains is an algebraic system supporting no propagating wave~\cite{Mak_FDTD}. The construction proposed in the preceding work~\cite{Mak26_2D} avoids that by separating the state, which integrates and is never drained, from the encoder, which converts the state to a symbol and carries its own error registers, so that the quantisation error is shaped towards the Nyquist frequency where the grid's dispersion relation guarantees no physical mode lives~\cite{Cal15, Contadini2003}.

Previous work~\cite{Mak26_2D} established that the scheme propagates waves on a two-dimensional (2D) lattice and converges with an error scaling as $\CFL^{0.93}$ in the Courant number for a general linear, homogeneous and lossless medium bounded by computationally inexpensive perfect electric conductor (PEC) boundaries. The present work addresses a more demanding question:~whether two-bit field variables remain viable for nonlinear electromagnetic simulations in complex, inhomogeneous media with absorbing boundary conditions, as required in many practical applications~\cite{Boyd, Taflove, Dis10, Mak11}. Specifically, can a ternary field support nonlinear optics? The observable is no longer simply a wave amplitude, but a frequency-conversion efficiency that the solver must generate dynamically, at an amplitude orders of magnitude below that of the pump. Moreover, the medium is inhomogeneous, so the update coefficients are no longer restricted to powers of two. Since nonlinear FDTD using floating-point arithmetic is well established~\cite{Zio03, Mak11}, it provides a direct reference for assessing the accuracy of the proposed approach.

I answer this question in the affirmative and show that the nonlinear problem exhibits a qualitatively different error structure from its linear counterpart. Section~\ref{sec:3d} formulates the linear scheme in three dimensions (3D), establishing the framework from which the benchmark one-dimensional (1D) problem studied in this work is derived. Section~\ref{sec:nomul} introduces the inhomogeneous and Kerr extensions, while Sections~\ref{sec:impl} and~\ref{sec:results} present the hardware implementation and experimental results, respectively. The complementary CPU and FPGA codes accompany the submission~\cite{Maksymov2026TernaryFDTD_NLO}.

\section{The ternary scheme in three dimensions}
\label{sec:3d}
In this section, I formulate a general linear Yee scheme in 3D, which establishes the framework from which the benchmark 1D problem is derived. I also explain why the ternary formulation requires no multiplication operations. 

\subsection{The Yee scheme}
In units $c=\mu_{r}=1$, Maxwell's curl equations read
\begin{equation}
  \frac{\partial\mathbf{H}}{\partial t}=-\nabla\times\mathbf{E},
  \qquad
  \varepsilon\,\frac{\partial\mathbf{E}}{\partial t}=\nabla\times\mathbf{H},
  \label{eq:maxwell}
\end{equation}
and are discretised on the staggered Yee grid~\cite{Yee66} with
$\Delta x=\Delta y=\Delta z=h$ and the Courant number
$\CFL=c\,\Delta t/h$ (Fig.~\ref{fig:concept}a). Writing the forward and backward differences along
$\xi\in\{x,y,z\}$ as
\begin{equation}
  (\dpl_{\xi}f)|_{\xi}=f|_{\xi+1}-f|_{\xi}\,,
  \qquad
  (\dmi_{\xi}f)|_{\xi}=f|_{\xi}-f|_{\xi-1}\,,
  \label{eq:diffops}
\end{equation}
with the half-integer Yee offsets understood and suppressed, the leapfrog update~\cite{Taflove} is
\begin{align}
  \Hx &\leftarrow \Hx + \CFL\bigl(\dpl_{z}\Ey-\dpl_{y}\Ez\bigr)\,,
  \label{eq:yee3d-hx}\\
  \Hy &\leftarrow \Hy + \CFL\bigl(\dpl_{x}\Ez-\dpl_{z}\Ex\bigr)\,,\\
  \Hz &\leftarrow \Hz + \CFL\bigl(\dpl_{y}\Ex-\dpl_{x}\Ey\bigr)\,,\\[2pt]
  \Ex &\leftarrow \Ex + \frac{\CFL}{\varepsilon}
        \bigl(\dmi_{y}\Hz-\dmi_{z}\Hy\bigr)\,,\\
  \Ey &\leftarrow \Ey + \frac{\CFL}{\varepsilon}
        \bigl(\dmi_{z}\Hx-\dmi_{x}\Hz\bigr)\,,\\
  \Ez &\leftarrow \Ez + \frac{\CFL}{\varepsilon}
        \bigl(\dmi_{x}\Hy-\dmi_{y}\Hx\bigr)\,,
  \label{eq:yee3d-ez}
\end{align}
the magnetic half-step preceding the electric one. Equations
\eqref{eq:yee3d-hx}--\eqref{eq:yee3d-ez} in floating-point arithmetic define the reference scheme throughout.

\subsection{The ternary updates and the encoder}
\label{sec:encoder}
A symbol $s\in\{-1,0,+1\}$ represents the field value $sA$, with $A$ a full-scale amplitude exceeding the largest field reached anywhere. An integer state $a\in\mathbb{Z}$ represents $aA/M$, with $M$ fixed-point units per symbol. The Courant number enters through the integer
\begin{equation}
  M_{C}=M\CFL\,,
  \label{eq:mc}
\end{equation}
which is required to be exact and which is most useful when $M$ and $\CFL$ are both powers of two~\cite{Mak26_2D}. Replacing every field in Eqs.~\eqref{eq:yee3d-hx}--\eqref{eq:yee3d-ez} by the neighbouring symbols (Fig.~\ref{fig:concept}a) gives
\begin{align}
  a^{H_{x}} &\!\leftarrow\! a^{H_{x}}
      + M_{C}\bigl(\dpl_{z}s^{E_{y}}-\dpl_{y}s^{E_{z}}\bigr)\,,
  \label{eq:tern3d-hx}\\
  a^{H_{y}} &\!\leftarrow\! a^{H_{y}}
      + M_{C}\bigl(\dpl_{x}s^{E_{z}}-\dpl_{z}s^{E_{x}}\bigr)\,,\\
  a^{H_{z}} &\!\leftarrow\! a^{H_{z}}
      + M_{C}\bigl(\dpl_{y}s^{E_{x}}-\dpl_{x}s^{E_{y}}\bigr)\,,\\[2pt]
  a^{E_{x}} &\!\leftarrow\! a^{E_{x}}
      + C\bigl(\dmi_{y}s^{H_{z}}-\dmi_{z}s^{H_{y}}\bigr)\,,\\
  a^{E_{y}} &\!\leftarrow\! a^{E_{y}}
      + C\bigl(\dmi_{z}s^{H_{x}}-\dmi_{x}s^{H_{z}}\bigr)\,,\\
  a^{E_{z}} &\!\leftarrow\! a^{E_{z}}
      + C\bigl(\dmi_{x}s^{H_{y}}-\dmi_{y}s^{H_{x}}\bigr)\,,
  \label{eq:tern3d-ez}
\end{align}
with
\begin{equation}  C=\mathrm{round}\!\left(M_{C}/\varepsilon_{\mathrm{eff}}\right)
  \label{eq:coef}
\end{equation}
being the only coefficient describing the properties of the material. The amplitude $A$ cancels identically from Eqs.~\eqref{eq:tern3d-hx}--\eqref{eq:tern3d-ez}, surviving only in the source injection and the reconstruction, and every operation is an integer addition or a comparison.

The state is converted to a symbol by an error-feedback modulator with two registers (Fig.~\ref{fig:concept}b),
\begin{align}
  v   &= x+\sigma\bigl(3e_{1}-e_{2},\,1\bigr)\,, \nonumber\\
  s   &= \operatorname{clamp}\bigl(\lfloor v/M\rceil,-1,+1\bigr)\,,
  \label{eq:enc}\\
  e_{2}&\leftarrow e_{1}\,,\qquad
  e_{1}\leftarrow\operatorname{clamp}\bigl(v-sM,-2M,+2M\bigr)\,,
  \nonumber
\end{align}
where $\sigma(v,n)$ denotes a symmetrically rounded right shift by $n$ places,
\begin{equation}
  \sigma(v,n)=\operatorname{sgn}(v)\,
  \Bigl\lfloor\bigl(|v|+2^{\,n-1}\bigr)/2^{\,n}\Bigr\rfloor\,,
  \label{eq:srs}
\end{equation}
so that $\sigma(-v,n)=-\sigma(v,n)$ exactly. One pair of registers is carried per component per cell, realising the noise transfer function~\cite{Cal15}
\begin{equation}
  \Nz=(1-z^{-1})\bigl(1-\tfrac12 z^{-1}\bigr)\,,
  \label{eq:ntf}
\end{equation}
whose zero at $z=1$ shapes the quantisation error away from low
frequency. The final clamp is an anti-windup bound that keeps the encoder state bounded by construction. Every operation in
Eq.~\eqref{eq:enc} is an addition, a constant shift or a select:~$3e_{1}$ is $(e_{1}\!\ll\!1)+e_{1}$, the division by two is the rounded shift $\sigma(\cdot,1)$ of Eq.~\eqref{eq:srs}, the division by $M$ is an arithmetic shift and $sM$ is a three-way select. The rounding in Eq.~\eqref{eq:srs} is load-bearing rather than a detail of implementation. A bare arithmetic right shift floors towards minus infinity, so it maps an odd $3e_{1}-e_{2}$ and its negation to feedback values differing by one fixed-point unit, breaking the odd symmetry $s(-x)=-s(x)$ of the encoder. The consequence is not a loss of accuracy but a loss of a symmetry the physics possesses, and it is measured in Sec.~\ref{sec:forbid}. The realised feedback coefficients differ from exactly $3/2$ and $-1/2$ by at most one fixed-point unit and, because the reference implementation evaluates the identical integer operator, the two symbol streams agree bit for bit rather than merely closely.

\subsection{Why no multiplication is required}
Each bracket in Eqs.~\eqref{eq:tern3d-hx}--\eqref{eq:tern3d-ez} is the difference between two first differences taken along different axes. The four ternary operands are separate state variables read at the same half-step, with no instantaneous constraint coupling their values; they can therefore vary independently over the alphabet. Writing the bracket as $d=(s_{1}-s_{2})-(s_{3}-s_{4})$, where each $s_i\in\{-1,0,+1\}$, each inner difference spans $\{-2,\ldots,+2\}$. The outer difference consequently spans
\begin{equation}
  d\in\{-4,-3,-2,-1,0,+1,+2,+3,+4\}\,,
  \label{eq:drange}
\end{equation}
with every value attainable; for example, $d=3$ for $(s_1,s_2,s_3,s_4)=(+1,-1,0,+1)$. These nine values exhaust the possible bracket outputs and thus define the complete operand alphabet required by the hardware.

Because $d$ is an integer of magnitude at most four, the products $M_{C}d$ and $Cd$ are selections among $\{0,\pm C,\pm 2C,\pm 3C,\pm 4C\}$: the sign is absorbed by the add--subtract that follows, $2C$ and $4C$ are left shifts by one and two places and $3C$ is stored beside $C$ in the same ROM word, so no arithmetic is needed to form it (Fig.~\ref{fig:concept}b). A nine-way selection network thus replaces the general multiplier for any representable permittivity, including the field-dependent $\varepsilon_{\mathrm{eff}}$ of a Kerr medium (Fig.~\ref{fig:concept}c). The coefficient may vary from cell to cell and from step to step; the operand alphabet does not. Inhomogeneity and nonlinearity therefore change only the contents of the table, never the width of the selection, and the cost of the update is independent of the medium. The reduction does not come from the medium being simple:~it comes from the operand alphabet being two bits wide.

\subsection{The spectral gap}
The ternary-valued FDTD method relies on a spectral separation between the band the physics occupies and the band into which the encoder pushes its error~\cite{Mak26_2D}. For a leapfrog discretisation whose spatial operator has largest eigenvalue $\rho$, the highest supported temporal frequency, in cycles per step, is
\begin{equation}
  f_{\max}=\pi^{-1}\arcsin\!\bigl(\CFL\sqrt{\rho}\bigr)\,,
  \label{eq:fmax}
\end{equation}
with $\rho=3$ on the homogeneous 3D Yee grid, against $\rho=2$ in 2D and $\rho=1$ in 1D. Since $\Nz$ vanishes at $f=0$ and is largest at the Nyquist frequency $1/2$, the oversampling ratio is
\begin{equation}
  \mathrm{OSR}=\frac{1/2}{f_{\max}}\simeq\frac{\pi}{2\CFL\sqrt{\rho}}
  \;=\;\frac{0.91}{\CFL},\;\frac{1.11}{\CFL},\;\frac{1.57}{\CFL}
  \label{eq:osr}
\end{equation}
in 3D, 2D and 1D, respectively. The spectral separation is therefore reduced in 3D relative to 2D and improved in 1D, but is not qualitatively altered:~$\mathrm{OSR}$ grows without bound as $\CFL\rightarrow0$ in every case and the $\sqrt{\rho}$ dependence is weak. Algorithm~\ref{alg:step3d} gives one complete time step in 3D:~the six components each carry a state accumulator and an error-register pair, so the per-cell storage accounting is per component.
\begin{alg}[t]
\algcaption{One ternary time step in 3D. $(\alpha,\beta,\gamma)$ runs over the cyclic permutations of $(x,y,z)$, which reproduces Eqs.~\eqref{eq:tern3d-hx}--\eqref{eq:tern3d-ez} exactly.}\label{alg:step3d}
\begin{algorithmic}[1]
\Function{Encode}{$x$; $e_{1},e_{2}$ \textbf{by reference}}
  \State $v \gets x + \sigma(3e_{1}-e_{2},1)$
         \Comment{rounded, not floored}
  \State $s \gets \operatorname{round}(v/M)$ clamped to $[-1,+1]$
  \State $e_{2} \gets e_{1}$; \quad
         $e_{1} \gets \operatorname{clamp}(v-sM,-2M,2M)$
         \Comment{anti-windup}
  \State \Return $s$
\EndFunction
\Statex
\ForAll{$\mathbf{H}$ cells, $(\alpha,\beta,\gamma)$ cyclic}
  \State $a^{H_{\alpha}} \mathrel{+}= M_{C}\,
         (\dpl_{\gamma}s^{E_{\beta}}-\dpl_{\beta}s^{E_{\gamma}})$
  \State $s^{H_{\alpha}} \gets \Call{Encode}{a^{H_{\alpha}};\,
         e^{H_{\alpha}}_{1},\,e^{H_{\alpha}}_{2}}$
\EndFor
\Statex
\ForAll{interior $\mathbf{E}$ cells, $(\alpha,\beta,\gamma)$ cyclic}
  \State $C \gets \textsc{Rom}\bigl[\mathrm{mat},\,
         |a^{E_{\alpha}}|\gg(\log_{2}\!M-n_{\mathrm{NL}})\bigr]$
  \State $a^{E_{\alpha}} \mathrel{+}= C\,
         (\dmi_{\beta}s^{H_{\gamma}}-\dmi_{\gamma}s^{H_{\beta}})$
  \State $s^{E_{\alpha}} \gets \Call{Encode}{a^{E_{\alpha}};\,
         e^{E_{\alpha}}_{1},\,e^{E_{\alpha}}_{2}}$
\EndFor
\Statex
\State inject the source into $a^{E}$ with its rounding residual
\State apply the boundary condition to $s^{E}$, $a^{E}$ \emph{and}
       $e^{E}_{1},e^{E}_{2}$ alike
\end{algorithmic}
\end{alg}

\subsection{Reduction to one dimension}
\label{sec:1dred}
Let the structure and the fields be invariant along $y$ and $z$, so that $\dpl_{y}=\dpl_{z}=\dmi_{y}=\dmi_{z}=0$. The six coupled update Eqs.~\eqref{eq:tern3d-hx}--\eqref{eq:tern3d-ez} then separate into two mutually independent transverse pairs, $(\Ez,\Hy)$ and $(\Ey,\Hz)$, each propagating along $x$. Taking the first,
\begin{align}
  \Hy &\leftarrow \Hy + \CFL\,\dpl_{x}\Ez\,,
  \label{eq:tern1d-h}\\
  \Ez &\leftarrow \Ez + \frac{\CFL}{\varepsilon_{\mathrm{eff}}}\,
                        \dmi_{x}\Hy\,,
  \label{eq:tern1d-e}
\end{align}
and the second pair is identical up to a sign. In an isotropic medium the two are uncoupled, so a single pair carries the whole problem (coupling between them arises only through the anisotropy of $\chi^{(3)}$, which is outside the scope of the present work~\cite{Dis10}).

The reduction is not merely a smaller problem but a different one for the arithmetic. Each bracket in Eqs.~\eqref{eq:tern3d-hx}--\eqref{eq:tern3d-ez} is a difference of two first differences and spans nine values, Eq.~\eqref{eq:drange}. Equations~\eqref{eq:tern1d-h}--\eqref{eq:tern1d-e} contain a single first difference, so the multiplicand is confined to
\begin{equation}
  d = s_{i}-s_{i-1} \in \{-2,-1,0,+1,+2\}\,,
  \label{eq:fiveway}
\end{equation}
and the selection narrows from nine ways to five, every member of which is a shift. The stability limit relaxes from
$\CFL\le1/\sqrt{3}$ to $\CFL\le1$ and, as per Eq.~\eqref{eq:osr}, the oversampling ratio becomes $1.57/\CFL$ rather than $0.91/\CFL$.

Three considerations make this reduction the right place to establish the scheme for nonlinear optics. The first is physical~\cite{Dis10, Taflove, Mak11}. Third-order processes are weak per unit length, so an appreciable conversion requires an interaction length of many wavelengths, and in a guided geometry the transverse field profile is fixed by the mode while the nonlinear dynamics unfold along the propagation axis alone. A longitudinal model is therefore the natural one and it is the model that practice adopts. For instance, Dissanayake~\emph{et al.}\ formulate an anisotropic $\chi^{(3)}$ FDTD for silicon waveguides in 3D and then reduce it, for the numerical work, to a pair of 1D FDTD models along the guide, with which they treat self- and cross-phase modulation, two-photon absorption and stimulated Raman scattering~\cite{Dis10}. (Similar reduction strategies remain applicable to nonlinear acoustics~\cite{Mak_FDTD}.) 

The second is computational. A one-dimensional domain admits a single propagation direction, a single polarisation and a single interface, so there are few mechanisms besides the arithmetic that could break the discrete symmetry on which the test of Sec.~\ref{sec:forbid} rests. (In 3D, numerical anisotropy, polarisation coupling and the edges and corners of the domain would each have to be excluded before an anomaly in a forbidden bin could be attributed to quantisation.) Cost reinforces the choice:~the characterisation rests not on one run but on an ensemble---a Courant sweep, twelve analysis windows at each point and controls at $\beta=0$ and at several values of $M$---and the cell count of a domain $L$ long grows as $(L/\Delta x)^{D}$ at the same Courant number and resolution. Decisively, the 1D configuration fits the programmable logic of a small device, so the same arithmetic can be run bit-exactly on hardware. (A 3D domain of comparable extent cannot and the claim could not be tested on a small FPGA at all.)

The third is methodological. One dimension admits a reference free correctness check that higher dimensions make no easier:~a centrosymmetric $\chi^{(3)}$ generates only odd harmonics, so every even bin of the measured spectrum is forbidden by symmetry and its occupancy measures the quantisation floor directly, without recourse to a reference solution (Table~\ref{tab:harm}). One dimension has likewise long served as the setting in which FDTD schemes are first stated and tested~\cite{Mak11}, and from which the method has been carried into acoustics and the time-dependent Schr\"odinger equation~\cite{Taflove,Sullivan}.

What the reduction omits should be stated plainly. Diffraction, transverse walk-off, vectorial and anisotropic nonlinear response and polarisation coupling are all absent, and phase matching degenerates to a condition on dispersion alone. None of these, however, bears on the question this paper asks. The quantities measured below---the agreement of the ternary and floating-point harmonic ratios, the quantisation floor, the existence of an optimal Courant number---are properties of the arithmetic rather than of the dimensionality.

\section{Inhomogeneous and nonlinear media}
\label{sec:nomul}
The argument in Sec.~\ref{sec:3d} removes the multiplier by making one factor of the product small:~the operand $d$ is a difference of ternary symbols, so it can take only five values in 1D and nine in 3D. The product can therefore be replaced by a selection among shifted copies of the coefficient. However, this argument says nothing about the other factor. It does not explain how the coefficient $C$ is obtained and assuming that $C$ is a single compile-time constant effectively restricts the scheme to vacuum. In nonlinear optics, $C$ is not constant:~it varies from cell to cell in a structured medium and, in a Kerr medium, changes at each time step with the field amplitude in the same cell. The apparent elimination of the multiplier may therefore simply reflect its relocation from the curl operation to the material coefficient, rather than its genuine removal.

This section addresses that concern. Four parts of the extended scheme appear to require a multiplier or a divider at run time:~the coefficient $C=M_{C}/\varepsilon$ for a medium of fixed contrast, the same coefficient when $\varepsilon$ depends on the local field, the reflection coefficient of the absorbing boundary and the phase advance
of the source. They are handled differently in the implementation, but they share a principle. A general multiplier is needed only when both factors are unknown at run time. If one factor instead comes from a small, enumerable set, remains fixed throughout the design's lifetime or can be made exact by a suitable choice of a free parameter, the multiplication can be resolved before the design is built rather than performed while it runs.

The following subsections demonstrate four ways to do this:~a select among precomputed multiples, where one factor is a small integer from a bounded range; table lookup for a coefficient that varies over a small set of values; canonical signed-digit expansion for a fixed coefficient that is rational but not dyadic, so that its binary expansion does not terminate; and an exact recurrence for a quantity that advances at every time step, where accumulated rounding error rather than instantaneous precision is the concern. In all four cases, the arithmetic is carried out by the reference program ahead of synthesis, leaving only a select, a lookup table, a few shifts or a counter in the hardware.

Although the operator \texttt{*} does not appear in the register-transfer description, in principle a synthesiser can still infer a hardware multiplier from a shift-and-add tree or a table-indexed product if it considers that implementation more efficient. To avoid this scenario, synthesis is performed with all DSP primitives disabled, i.e.~no dedicated hardware resources designed to perform multiplication operations efficiently are used~\cite{Fingeroff2010}.

\subsection{Dielectric contrast and the Kerr table}
For $\varepsilon_{r}=2.25$ with $M_{C}=2^{10}$, Eq.~\eqref{eq:coef} gives $C=\mathrm{round}(1024/2.25)=455$, neither a power of two nor a shift, yet Eq.~\eqref{eq:fiveway} absorbs it at the cost of one multiplexer. The realised permittivity is $M_{C}/C=2.250549$, an error of $0.024\%$.

The instantaneous Kerr effect admits several FDTD treatments, which differ not merely in accuracy but in how they interpret the physics of a permittivity that depends on the field being solved for:~the nonlinear term may be solved implicitly at each step by iteration~\cite{Taflove, Mak_FDTD}, the polarisation may be carried as an auxiliary state satisfying its own differential equation~\cite{Sullivan, Mak_FDTD} or the index may be evaluated explicitly from the field already held~\cite{Mak11}. I take the last and simplest of these~\cite{Mak11}:~$\varepsilon_{\mathrm{eff}}$ is formed from the state stored in the cell at the start of the step, so the update remains explicit and single-pass, at the cost of lagging the index by one time step---a lagging-index error of order $O(\Delta t)$ and therefore first order in $\CFL$. It is kept small by the oversampled operation of Eq.~\eqref{eq:osr}, it is the dominant contribution to the pump amplitude error measured in Sec.~\ref{sec:results} and it is entirely distinct from the quantisation-noise bias identified below.

The choice is deliberate but not load-bearing. Whichever treatment is adopted, its consequence for the arithmetic is the same~\cite{Mak11}:~a coefficient that varies from cell to cell and from step to step and must therefore be produced anew for each update. The argument of Sec.~\ref{sec:3d} is untouched by this, since it turns on the width of the operand rather than the provenance of the coefficient. A table, an iteration or an auxiliary state would each supply $C$; none of them would widen $d$ beyond two bits. 

For an instantaneous Kerr medium, $\varepsilon_{\mathrm{eff}}=\varepsilon_{r}(1+\beta E^{2}/A^{2})$ and squaring the $18$-bit state would be a genuine multiplication. Instead the coefficient is read from a table addressed by the material index and the top $n_{\mathrm{NL}}=6$ bits of $|a|$,
\begin{align}
  C      &= \mathrm{ROM}\bigl[\,\mathrm{mat}_{i}\,\|\,\iota\,\bigr]\,,
  \nonumber\\
  \iota  &= \min\bigl(|a|\gg(m-n_{\mathrm{NL}}),\;
                      2^{n_{\mathrm{NL}}}-1\bigr)\,,
  \label{eq:rom}
\end{align}
where $m=\log_{2}M$ and the operator $\|$ denotes bit concatenation, so that the table output feeds the same five-way select. For $\beta=2$ the table spans $C=455$ down to $153$ across its $64$ entries. The material index is derived from the cell address, $\mathrm{mat}_{i}=[\,i\ge x_{\mathrm{int}}\,]$, so the geometry occupies no memory and vacuum and dielectric share one datapath.

There are two important physical consequences of Eq.~\eqref{eq:rom} and they are closely related:~the first provides a way to measure the quantisation noise, while the second predicts how that noise affects the nonlinear response.

The address is the magnitude $|a|$, so $C$ is an even function of the state and the polarisation $\varepsilon_{\mathrm{eff}}E$ is odd. A memoryless odd nonlinearity driven at $f_{0}$ can produce a steady-state response satisfying $E(t+T/2)=-E(t)$. This half-wave antisymmetry excludes even Fourier harmonics, leaving only $f_{0},3f_{0},5f_{0},\dots$, consistent with the symmetry of a centrosymmetric medium with a third-order susceptibility $\chi^{(3)}$~\cite{Boyd}.

What matters here is that this symmetry is exact in the arithmetic, not merely approximate. It does not depend on the cancellation of nearly equal numbers or on the accuracy of the ROM entries. Even a coarse table cannot generate an even harmonic through this mechanism, because $+a$ and $-a$ always address the same ROM word. The order of operations is essential:~the magnitude is calculated before the shift that selects the top $n_{\mathrm{NL}}$ bits. Shifting first would round negative values towards the lower integer address, mapping $+a$ and $-a$ to addresses that differ by one least significant bit. This would break the symmetry and introduce even-order harmonics.

The even harmonics therefore provide a reference measure of the numerical noise floor. Table~\ref{tab:harm} and Sec~\ref{sec:forbid} use them for this purpose. 

The second property is that the ROM address depends on the instantaneous state, including the shaped quantisation noise carried by the field. Writing $a=a_{s}+n$, where $n$ is zero-mean and uncorrelated with the signal, gives $\langle a^{2}\rangle=\langle a_{s}^{2}\rangle+\sigma_{n}^{2}$. The ROM therefore sees a mean-square state larger than that of the signal alone, making the effective Kerr contribution slightly stronger, with a leading-order fractional bias of $\beta\sigma_{n}^{2}/M^{2}$. At $\CFL=1/256$, the measured full-band noise is $\sigma_{n}=7.6\times10^{-4}A$, giving a bias of order $10^{-6}$, negligible compared with the other errors in the scheme.

\subsection{Absorbing boundaries:~signed-digit shift--add}
A first-order Mur condition~\cite{Mur81}, $E^{n+1}_{0}=E^{n}_{1}+\kappa\,(E^{n+1}_{1}-E^{n}_{0})$ with
$\kappa=(\CFL-1)/(\CFL+1)$, has a coefficient that is rational but not dyadic. At $\CFL=2^{-8}$, for example, $\kappa=-255/257$. Since $257$ is prime and is not a power of two, the binary expansion does not terminate. The right-hand wall of a half space compounds the difficulty, since it lies inside the dielectric where the local Courant number is $\CFL/n_{r}$, so that its leading correction carries a factor $4/3$ for $n_{r}=3/2$ in place of $2$.

Both are handled by expanding $\kappa$ greedily into a signed-digit sum of at most five signed powers of two, which the reference program computes and emits as the expression the hardware evaluates. For $\CFL=2^{-8}$ the vacuum wall yields
\begin{equation}
  \kappa_{L}\rightarrow
  -\,\sigma(d,0)+\sigma(d,7)-\sigma(d,15)+\sigma(d,23)\,,
  \label{eq:csd}
\end{equation}
where $\sigma$ is the symmetrically rounded right shift of Eq.~\eqref{eq:srs}. These four terms are exactly the first four of the geometric expansion $\kappa=-(1-2\CFL+2\CFL^{2}-2\CFL^{3}+\cdots)$, because $2\CFL$, $2\CFL^{2}$ and $2\CFL^{3}$ are themselves powers of two at this Courant number. The greedy search recovers the series term by term rather than approximating it and the realised coefficient differs from the exact value by $4.6\times10^{-10}$, consistent to two significant figures with the first omitted term. At the dielectric wall, the leading correction $(4/3)\CFL$ is not a power of two and $4/3=1.0101\ldots_{2}$ contributes a geometric tail rather than a single digit. The expansion therefore needs five terms rather than four, yet leaves $6.8\times10^{-6}$, approximately four orders of magnitude greater than that of the vacuum wall.

The rounding in $\sigma$ is not cosmetic. A bare arithmetic shift floors toward minus infinity, so every application carries a mean error of $-\tfrac12$ least-significant bit and the boundary update accumulates that error step after step rather than averaging it away. Over the $8\times10^{5}$ steps of a run one such shift drifts by $1.5A$, already seven times the pump amplitude at the probe, and the expansion applies four or five of them per boundary cell per step. A boundary wandering by several times $A$ would not announce itself as a
rounding fault:~nothing overflows, no individual update looks
unreasonable and the only symptom is a field that leaves the physical range. Rounding the magnitude and restoring the sign makes the error zero-mean, leaving a random walk of order $\sqrt{N}$ least-significant bits, some $10^{-3}A$ over the same run, against the $1.5A$ of the systematic case.

\subsection{Source:~an exact prescaled counter}
The excitation source is based on a $2^{10}$-entry sine table. If the wavelength $\lambda$ is constrained to a power of two, the number of steps per source period, $\lambda/\CFL=2^{14}$, is also a power of two and the table index is an exact prescaled counter, i.e.~each entry is held for precisely $2^{4}$ steps and there is no phase accumulator.

The constraint is worth stating because the obvious alternative---a truncated phase accumulator, the textbook direct digital synthesiser~\cite{Kroupa}---fails in a way specific to the nonlinear problem. Phase truncation places a periodic sawtooth on the source amplitude, which in a Kerr medium mixes with the pump and fills the even-harmonic bins. I measured second-harmonic suppressions of $33.9$, $54.4$ and $35.4$~dB at table depths of $10$, $12$ and $14$~bits, non-monotonic in the depth because the truncation residual
$e_{n}=n\Delta\phi \bmod 2^{B}$, with $\Delta\phi$ the phase increment and $B$ the number of discarded low bits, is periodic with period $2^{B}/\gcd(\Delta\phi,2^{B})$, where $\gcd$ denotes the greatest common divisor. Whether a line of that period falls near $2f_{0}$ is an arithmetic accident of the increment.

The same expression gives the remedy. Restricting $\lambda$ to a power of two makes the number of steps per source period a power of two, so $\Delta\phi$ is one also, the greatest common divisor saturates at $\Delta\phi$ itself and the period collapses to $16$~steps---a divisor of the source period rather than a quantity incommensurate with it. Every line the residual can produce then lies at a multiple of $2^{k}f_{0}=1024f_{0}$, the rate at which the table advances, three orders of magnitude above the harmonics being measured. For comparison, $\lambda=60$ cells at the same Courant number gives a residual period of about $10^{6}$~steps, a comb of spacing $0.015f_{0}$ whose lines can fall anywhere, $2f_{0}$ included. The exact counter is therefore not an improved approximation to the accumulator:~with the steps per period a power of two, no phase need be accumulated at all, the table index being $(n\gg P)\,\&\,(2^{k}-1)$, a shift and a mask, exact for every $n$. There is nothing left to truncate.
\begin{figure*}[t]
\centering
\includegraphics[width=\textwidth]{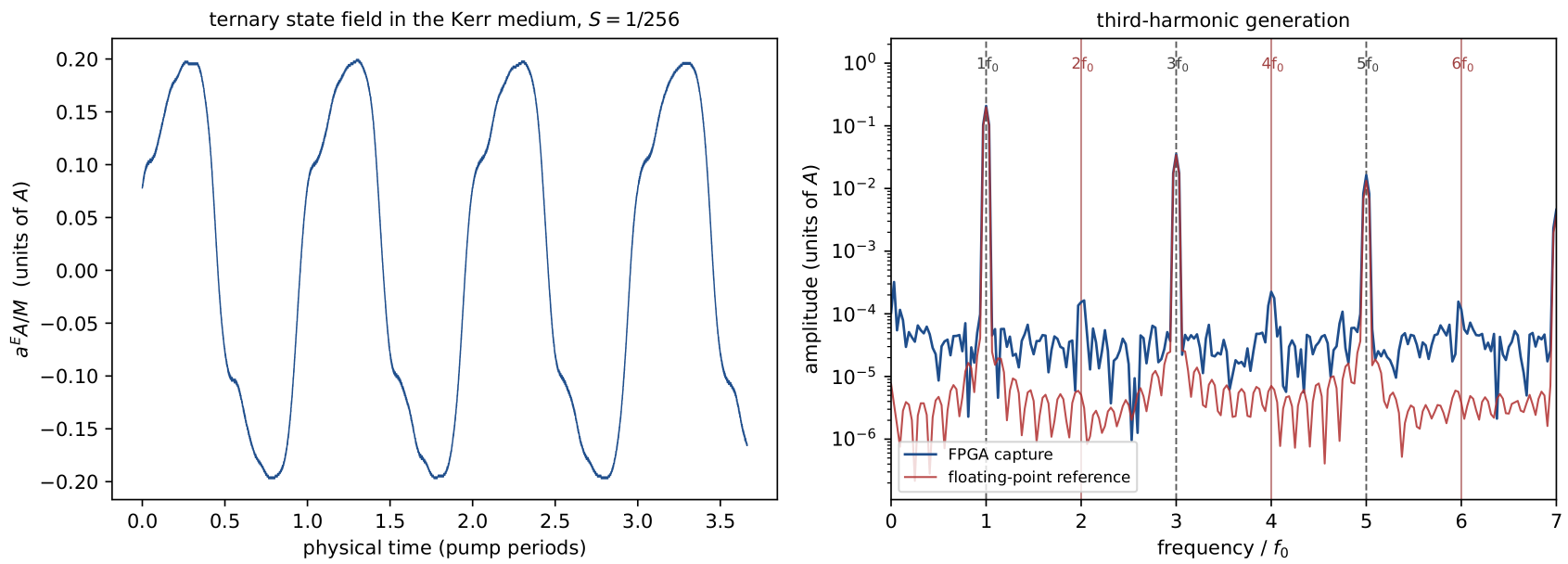}
\caption{\label{fig:spectrum}%
  Both panels are the hardware capture of Sec.~\ref{sec:results}\,D, at a probe $128$ cells inside the Kerr half space. (Left)~The ternary state field, plotted as $a^{E}A/M$ in units of $A$, over the first few pump periods at $\CFL=1/256$. The waveform is assembled from a stream of two-bit symbols, yet is visibly non-sinusoidal:~that distortion is due to the higher-order harmonics, which the right panel resolves. (Right)~Harmonic content at the same probe, against a floating-point reference computed on the same grid. The odd harmonics $3f_{0}$ and $5f_{0}$ are generated and agree with the reference. The even harmonics, which a centrosymmetric $\chi^{(3)}$ cannot produce, lie $90$~dB below the pump in the reference and sit at the ternary solver's quantisation floor. The gap between $3f_{0}$ and that floor is the figure-of-merit of the measurement.}
\end{figure*}

\section{Implementation and verification}
\label{sec:impl}
The code accompanying this work~\cite{Maksymov2026TernaryFDTD_NLO} is a single artefact serving two kinds of reader. The C programming language reference needs nothing but a compiler and runs anywhere; the register-transfer description targets an FPGA board costing a few tens of US dollars. These are not two implementations of one scheme, which would be free to drift apart, but one implementation with two front ends:~the C reference emits the parameter file, the coefficient table and the source table against which the hardware is compiled, and the hardware is then checked against the C reference step by step, bit for bit, with no tolerance anywhere in the comparison. That such a comparison is meaningful at all is a consequence of the scheme itself, rather than of the care taken in testing, since integer arithmetic has no rounding modes that could lead to disagreement.

Sharing a front end is what makes the comparison bit-exact, but it also means the floating-point model is not an independent implementation. An error in the discretisation itself, as distinct from an error in realising it, would appear in both and cancel between them. The forbidden even harmonics of Sec.~\ref{sec:forbid} close exactly that gap:~they follow from the symmetry of the susceptibility rather than from any particular discretisation, so they test the scheme against the physics instead of against a sibling program. The two checks are complementary rather than redundant and Sec.~\ref{sec:forbid} reports a defect that only the second could find.

Readers interested mainly in the algorithm or its software implementation can skip Sec.~\ref{sec:impl} without losing the thread of the paper. Nevertheless, this section is important because it tests the paper's central claim rather than simply stating it. Whether an update can be performed without a multiplier is ultimately a question of hardware implementation and source code alone cannot answer it. The decisive test is to synthesise the design into a netlist (i.e.~enumerated arrangement and interconnection of components) with dedicated multipliers disabled and check whether the arithmetic is implemented using shifts, selections and lookup tables, or whether multiplication has simply been concealed in the description. Section~\ref{sec:impl} presents the results of this test on an FPGA that provides 48 embedded multipliers but no dedicated floating-point hardware. The solver occupies $3285$ of the device's $20{,}736$ logic sites, $15.8\%$ of the part, and none of its multipliers of any width (Table~\ref{tab:util}), making a complete nonlinear optical solver accessible to teaching laboratories and embedded instruments (Sec.~\ref{sec:onboard}).

\subsection{Core}
The solver is implemented on a Tang Nano 20K (Gowin GW2AR-18C~\cite{sipeed_tang_nano_20k}:~$20{,}736$ LUT4, $15{,}552$ flip-flops, $828$~Kbit of block SRAM in $46$ blocks, $48$ embedded $18\times18$ multipliers, $64$~Mbit of SDRAM) in a single clock domain driven directly by the $27$~MHz onboard oscillator. Neither of the device's two phase-locked loops is used and none of its $48$ multipliers:~the design declines both, rather than lacking them. Nine of the $46$ block memories are occupied:~eight hold the two symbol planes and, for each of $E$ and $H$, the state and the encoder's two error registers;~the ninth holds the coefficient and source tables, which are read-only. Addresses are bit concatenations, which requires the cell count to be a power of two and removes the last place a multiplier could hide.

Table~\ref{tab:util} reports the resource usage after place and route. Every multiplier-bearing primitive the architecture offers is unused:~not only the $48$ $18\times18$ multipliers but the $9\times9$ and $36\times36$ variants, the fused multiply--accumulate blocks and the $54$-bit arithmetic units, all of which would have been available to the synthesiser had it chosen to infer one. The critical path, $7.93$~ns, runs from the clocked output of the source sine table through the carry chain of the source accumulator;~$2.26$~ns of it, $29\%$, is the block-memory clock-to-output delay. The limiting path therefore lies in the excitation rather than in the ternary arithmetic, which is consistent with a datapath whose widest operation is a five-way select followed by an addition.

Memories have a registered address and a registered output, so data for an address presented in cycle $n$ are available in cycle $n+2$, and each sweep is written out state by state around that latency. A cycle-exact replay of the state machine counts $7167$ cycles per time step, or $265~\mu$s at $27$~MHz. One step is reported to the host as one $6$-byte frame and the solver stalls until that frame has been transmitted, so one received frame is one time step by construction. At $115200$~baud the frame costs $521~\mu$s and the link, not the solver, sets the rate.

\begin{table}[t]
\caption{\label{tab:util}%
  Resource usage of the complete solver after place and route on the
  Gowin GW2AR-18C. On this architecture, a carry-chain (ALU) cell occupies a lookup-table site, so the two rows share the $20{,}736$ available sites and the logic total is their sum. $F_{\max}$ is a conservative bound:~post-place-and-route timing analysis shows the design sustains at least this clock rate. In this work, the device is operated at $27$~MHz directly from the onboard oscillator and no multiplier-bearing primitive of any width is instantiated.}
\begin{ruledtabular}
\begin{tabular}{lrrr}
 resource & used & available & \% \\
\colrule
 LUT4                          & $2075$ & $20{,}736$ & $10.0$ \\
 ALU (carry chain)             & $1210$ & $15{,}552$ & $7.8$  \\
 \quad logic sites, total      & $3285$ & $20{,}736$ & $15.8$ \\
 flip-flops                    & $1057$ & $15{,}552$ & $6.8$  \\
 block SRAM (18~Kbit)          & $9$    & $46$       & $19.6$ \\
 $18\times18$ multipliers      & $0$    & $48$       & $0$    \\
 all other DSP primitives      & $0$    & $348$      & $0$    \\
 phase-locked loops            & $0$    & $2$        & $0$    \\
 I/O pins                      & $5$    & $384$      & $1.3$  \\
\colrule
 $F_{\max}$ & \multicolumn{3}{r}{$126.2$~MHz (critical path $7.93$~ns)} \\
\end{tabular}
\end{ruledtabular}
\end{table}

\subsection{Verification}
Four layers were used, in increasing scope. (i)~The encoder was checked against $19683$ exhaustively generated vectors covering both signs, the saturation boundary and the anti windup boundary, bit-exactly. (ii)~A cycle-accurate replay of the state machine, including the two-cycle memory latency, was compared against the C programming language reference step by step. (iii)~The register-transfer description was compared against the same reference, bit-exactly and with no tolerance. This layer is a test of the arithmetic rather than of the physics, so it is run on a reduced configuration ($128$ cells, $\lambda=16$, $\CFL=1/16$) chosen so that a short simulation still exercises every path:~the wave crosses the vacuum--dielectric interface at step $384$ and reaches the dielectric probe at step $640$, leaving some nine pump periods of propagation inside the Kerr medium, over which the coefficient table is addressed across $45\%$ of its range and the signed-digit Mur chains run at a non-dyadic coefficient, $\kappa_{L}=-15/17$. (iv)~The serial framing was decoded as the host decodes it, asserting one frame per step with a contiguous step counter.

A single generated parameter file, written by the reference
implementation, supplies every constant, including the boundary expressions Eq.~\eqref{eq:csd} themselves and the table contents, to the hardware, the simulation and the analysis alike, so the three cannot hold different opinions about any number. The complete source is available in Ref.~\cite{Maksymov2026TernaryFDTD_NLO}.

The same source was also synthesised for a Xilinx~XC7Z020 with all DSP primitives disabled. No multiplier-bearing primitive is inferred there either and the two technology mappers, which share no mapping code, lookup-table width or carry structure, infer the same nine 18\,Kbit block memories and agree on the number of carry bits to within 13\%. Register and lookup-table counts differ by tens of percent, as expected of quantities set by each tool's optimisation and packing rather than by the design.

This means that the multiplier-free property belongs to the arithmetic rather than to one vendor's technology mapper. A datapath can be free of multipliers under one mapper and not under another, since a shift-and-add tree or a table-indexed product may be absorbed into a hard multiplier by a tool that judges it cheaper. Therefore, two independent mappers declining to do so indicates that there is no multiplication in the datapath to find. What the two agree on is what the source fixes---the arrays, which become nine block memories in either technology, and the adders, whose carry bits are counted alike by a one-bit chain and a four-bit one. The scheme therefore ports, which the extrapolation of Sec.~\ref{sec:onboard} relies on:~the radiation-tolerant device considered there carries a third vendor's fabric and a third mapper, and the argument made for it assumes only that the datapath contains no multiplication.

\section{Results}
\label{sec:results}

\subsection{Third-harmonic generation}
The test problem, sketched in Fig.~\ref{fig:concept}c, is a $512$-cell domain, vacuum for $x<256$ and a Kerr dielectric medium with $\varepsilon_{r}=2.25$ and $\beta=2$ beyond it, with first-order Mur boundaries at both ends and a source of amplitude $A/2$ at cell $32$ driving at $\lambda=64$ cells. The record is $8\times10^{5}$ steps, or $48.8$ pump periods, of which the wave's arrival and one full domain transit are discarded before analysis. The value $\beta=2$ is chosen as an arithmetic stress test rather than as a model of a particular material. At the field amplitudes reached it produces a peak index shift of $\beta(E/A)^{2}\approx9\%$ of $\varepsilon_{r}$, strong for a Kerr medium but placing the third harmonic well above the quantisation floor, so that the floor itself can be measured. While the physical principles remain valid, the claims below concern the arithmetic properties of the computational scheme rather than the practical realism of the medium.

With $\lambda=64$ cells in vacuum and $n=\sqrt{\varepsilon_{r}}=1.5$, the grid carries $42.7$ cells per wavelength at the pump inside the dielectric, $14.2$ at the third harmonic and $8.5$ at the fifth. Both ternary and reference solvers run on the same grid, so numerical dispersion is common to them and cancels in the ratios reported below. It does, however, limit the accuracy with which the fifth harmonic is represented in either and the absolute value of $a_{5}$ should be read with that in mind.

Table~\ref{tab:harm} gives the harmonic content at a probe $128$ cells inside the dielectric at $\CFL=1/256$. Figure~\ref{fig:spectrum} shows the state field at that probe and the spectrum from which those amplitudes are taken. The third-harmonic conversion efficiency, here the amplitude ratio $\eta\equiv a_{3}/a_{1}$, is $17.02\%$ against $17.26\%$ for the floating-point reference, a ratio of $0.9863$. (The fifth harmonic, produced by cascaded four-wave mixing between the pump and the generated third harmonic~\cite{Mak13}, is also present and reproduced to within 17\%, the larger discrepancy reflecting the sensitivity of the higher cascade order to the pump amplitude error~\cite{Mak11}.) The pump amplitude carries a $+4.94\%$ excess. This is the lagging-index error of Sec.~\ref{sec:nomul}:~it is first order in $\CFL$ and falls with it, as Table~\ref{tab:sweep} confirms. It is not the quantisation-noise bias $\beta\sigma_{n}^{2}/M^{2}$ estimated in the same section, which is of order $10^{-6}$ and therefore more than four orders of magnitude too small to account for it.

\begin{table}[b]
\caption{\label{tab:harm}%
 Harmonic amplitudes $a_{k}$ at the dielectric probe, $\CFL=1/256$, in units of $A$, over $32$ whole source periods, ensuring that every harmonic falls on a bin. A centrosymmetric $\chi^{(3)}$ allows odd $k$ and forbids even $k$, and the coefficient table, addressed by $|a|$, remains exactly even regardless of its coarseness (Sec.~\ref{sec:nomul}). The even columns therefore quantify the solver's departure from odd symmetry rather than the table's accuracy:~for the ternary solver its quantisation floor together with the DC offset of the undrained accumulator and for the reference the residual leakage of the finite analysis window. The dB column gives the per-bin suppression of each line below the pump; the broadband floor quoted in the text---the worst of the three even bands, each integrated over a width $f_{0}/2$---is $3.4\times10^{-4}$, against $1.5\times10^{-4}$ in the single bin at $2f_{0}$.}
\begin{ruledtabular}
\begin{tabular}{crrrr}
 & \multicolumn{2}{c}{ternary} & \multicolumn{2}{c}{floating point} \\
 $k$ & $a_{k}$ & dB & $a_{k}$ & dB \\
\colrule
 1 & $2.07\!\times\!10^{-1}$ & ---  & $1.97\!\times\!10^{-1}$ & ---  \\
 2 & $1.54\!\times\!10^{-4}$ & 62.6 & $5.1\!\times\!10^{-6}$  & 91.8 \\
 3 & $3.52\!\times\!10^{-2}$ & 15.4 & $3.40\!\times\!10^{-2}$ & 15.3 \\
 4 & $2.24\!\times\!10^{-4}$ & 59.3 & $7.1\!\times\!10^{-6}$  & 88.9 \\
 5 & $1.63\!\times\!10^{-2}$ & 22.1 & $1.33\!\times\!10^{-2}$ & 23.4 \\
 6 & $1.17\!\times\!10^{-4}$ & 64.9 & $3.8\!\times\!10^{-6}$  & 94.3 \\
\end{tabular}
\end{ruledtabular}
\end{table}

\subsection{The forbidden harmonics as a self-check}
\label{sec:forbid}
In a centrosymmetric medium, every even-order susceptibility vanishes, so the lowest nonlinearity is $\chi^{(3)}$, which mixes three fields at a time~\cite{Boyd}. Each such interaction produces a signed combination $\pm f_{a}\pm f_{b}\pm f_{c}$ of frequencies already present~\cite{Mak13}. Starting from a pump at $f_{0}$, the first interaction reaches $3f_{0}$ and $f_{0}$ and, since a signed sum of three odd integers is odd, the odd multiples of $f_{0}$ are closed under the operation. No cascade of any order can therefore reach $2f_{0}$ or $4f_{0}$, while $5f_{0}$ follows from mixing the pump with the third harmonic it has generated.

The even bins are thus empty on physical grounds rather than merely small, which makes whatever occupies them a readout of the solver's own departure from that symmetry. The material model does not limit the reading:~the coefficient table is addressed by $|a|$, so $\pm a$ return the identical word and $\varepsilon_{\mathrm{eff}}$ is an exactly even function of the field (Sec.~\ref{sec:nomul}), whatever the accuracy of the table itself. What remains is the arithmetic, and the even bins measure it directly---down to the floor of the analysis, which the floating-point column of Table~\ref{tab:harm} calibrates.

In the floating-point reference the even bins lie $89$--$94$~dB below the pump. In the ternary solver they lie $59$--$65$~dB below it (Table~\ref{tab:harm}). The useful quantity is the margin of the real third harmonic above that level, integrated over a band of fixed physical width:~at $\CFL=1/256$ it is $40.3$~dB. Both levels are visible in the right-hand panel of Fig.~\ref{fig:spectrum}:~the even bins lie 90 dB below the pump in the reference trace and sit on a broadband floor in the ternary one.

What occupies the even bins is not quantisation noise alone. Any residual departure from exact odd symmetry appears there and in the present solver the dominant contribution is a small DC offset carried by the state accumulator. In fact, because the accumulator is never drained, a bias of even one unit per step does not average away but integrates and a DC component mixes with the pump through the Kerr term to place a deterministic line at $2f_{0}$. The even bins should therefore be read as a bound on the total departure from odd symmetry, of which the quantisation floor is one term.

This process is described in detail because it provides a useful means of self-verification. The truncated-phase source described in Sec.~\ref{sec:nomul} raised the even bins to within $34$~dB of the pump, while leaving the odd harmonics, pump amplitude and time-domain waveform unchanged within measurement accuracy. Furthermore, the encoder's feedback divide-by-two used a bare arithmetic right shift, which floors towards minus infinity and therefore violates $s(-x)=-s(x)$ by one state unit whenever $3e_{1}-e_{2}$ is odd. Integrated by the undrained accumulator and mixed with the pump through the Kerr term, this asymmetry reduced second-harmonic suppression by approximately $15$~dB. Since none of the built-in verification procedures for either the hardware or software implementation detected the discrepancy, it could only be exposed by testing a symmetry of the governing physics rather than agreement with a reference solution.
\begin{figure}[t]
\centering
\includegraphics[width=\columnwidth]{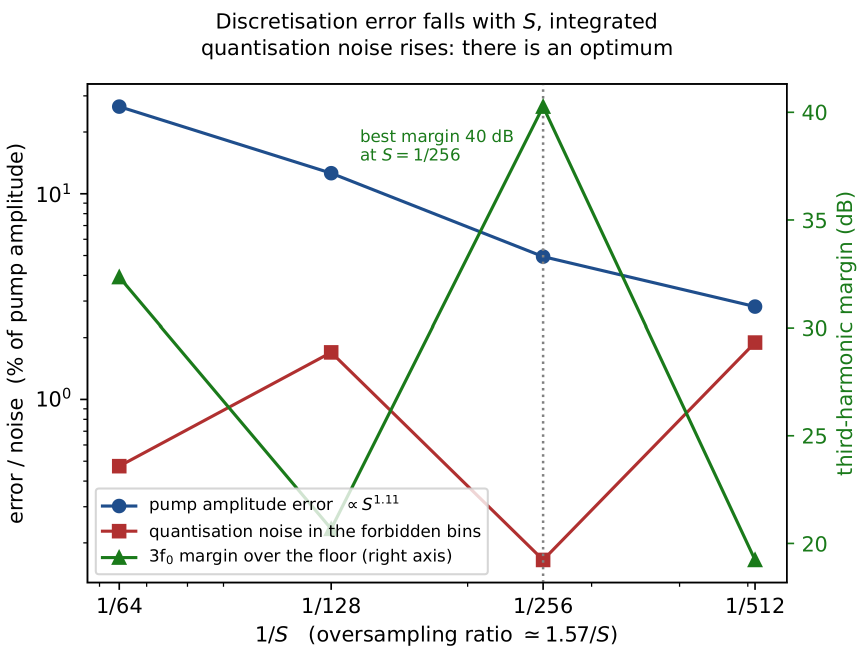}
\caption{\label{fig:convergence}%
  The trade-off:~the pump amplitude error falls as $\CFL^{1.11}$, while the quantisation noise integrated into the forbidden bins does not fall with it, since at fixed physical time $T$ the number of quantisation events to accumulate grows as $T/\CFL$. The two do not compete smoothly:~the noise is non-monotonic in $\CFL$, so the margin of the third harmonic over the floor is not a smooth single peak but is largest at $\CFL=1/256$ of the four values measured (Sec.~\ref{sec:optimum}).}
\end{figure}

\subsection{An optimal Courant number}
\label{sec:optimum}

Table~\ref{tab:sweep} repeats the measurement across a factor of eight in the Courant number at fixed physical time $T=N\CFL$, as required for a convergence study since holding the step count fixed would instead compare runs of different physical durations.

The pump amplitude error falls smoothly with the Courant number, as $\CFL^{1.11}$ over a factor of eight, but the margin of the third harmonic over the in-band quantisation noise does not follow it:~it is $32.3$, $20.7$, $40.3$ and $19.2$~dB at $\CFL=1/64$, $1/128$, $1/256$ and $1/512$, respectively. Figure~\ref{fig:convergence} plots the two quantities against $1/\CFL$. Re-measuring each point over 12 analysis windows (three record lengths and four sub-period offsets) changes no value by more than a decibel and leaves the ordering intact, so the variation between rows is a property of the solver and not of the measurement. A smooth competition between a falling discretisation error and a rising noise floor could not produce it:~$\CFL=1/128$ is worse than both its neighbours and the highest oversampling ratio in the sweep gives the worst margin of the four. The non-monotonicity is instead characteristic of the tonal behaviour of noise-shaped quantisers, whose in-band content depends on the signal amplitude relative to the quantiser step in a manner that is not smooth in the parameters~\cite{Widrow-Kollar}. The practical consequence is that the operating point is found by measurement rather than predicted:~$\CFL=1/256$ is the best of the four by $8.0$~dB, and I do not claim it is optimal for other choices of $M$,
$\lambda$ or drive amplitude.

This is absent from the linear problem~\cite{Mak26_2D}, where smaller $\CFL$ is simply better, and it is a design rule:~a ternary solver should be run at the oversampling ratio that maximises the margin of the physics of interest over the shaped noise, not at the smallest time step the hardware permits.
\begin{table}[b]
\caption{\label{tab:sweep}%
  Courant sweep at $M=2^{18}$, analysed over $32$ whole source periods at every point. The oversampling ratio is the 1D value $1.57/\CFL$ of Eq.~\eqref{eq:osr}. The pump amplitude error falls smoothly as $\CFL^{1.11}$, but the integrated in-band quantisation noise does not fall with it, and the third-harmonic margin over that noise is strongly non-monotonic. THG is the amplitude ratio $\eta=a_{3}/a_{1}$ and ratio is $\eta$ divided by its floating-point value on the same grid; both are computed from the unrounded amplitudes rather than from the three-figure entries of Table~\ref{tab:harm}.}
\begin{ruledtabular}
\begin{tabular}{lrrrrr}
 $\CFL$ & OSR & pump err & THG & ratio & margin \\
\colrule
 $1/64$  & 101 & $+26.6\%$ & $19.60\%$ & $1.135$ & $32.3$~dB \\
 $1/128$ & 201 & $+12.6\%$ & $18.33\%$ & $1.062$ & $20.7$~dB \\
 $1/256$ & 402 & $+4.94\%$ & $17.02\%$ & $0.986$ & $40.3$~dB \\
 $1/512$ & 804 & $+2.82\%$ & $17.33\%$ & $1.004$ & $19.2$~dB \\
\end{tabular}
\end{ruledtabular}
\end{table}

\subsection{Hardware measurement}
The same computation was then run on the FPGA and $8\times10^{5}$ consecutive time steps were captured with no lost, duplicated or reordered frames. The ternary column of Table~\ref{tab:harm} and the FPGA trace of Fig.~\ref{fig:spectrum} are that capture:~the harmonic content reported there was extracted from the hardware, not from the reference model, so its agreement with the floating-point column is a statement about the FPGA implementation rather than about the model. That the two implement the same integer recurrence exactly, rather than approximately, was established separately in simulation, where the register-transfer description and the C programming language reference were compared state by state over $3\times10^{3}$ steps of the reduced configuration of Sec.~\ref{sec:impl}\,B, with no tolerance and no mismatch:~neither a symbol nor a state differed at any step, and the comparison spans the arrival of the wave at the dielectric probe and nine subsequent pump periods within the nonlinear medium.

\section{Solving where the decision is made}
\label{sec:onboard}
\begin{figure*}[t]
\centering
\includegraphics[width=\textwidth]{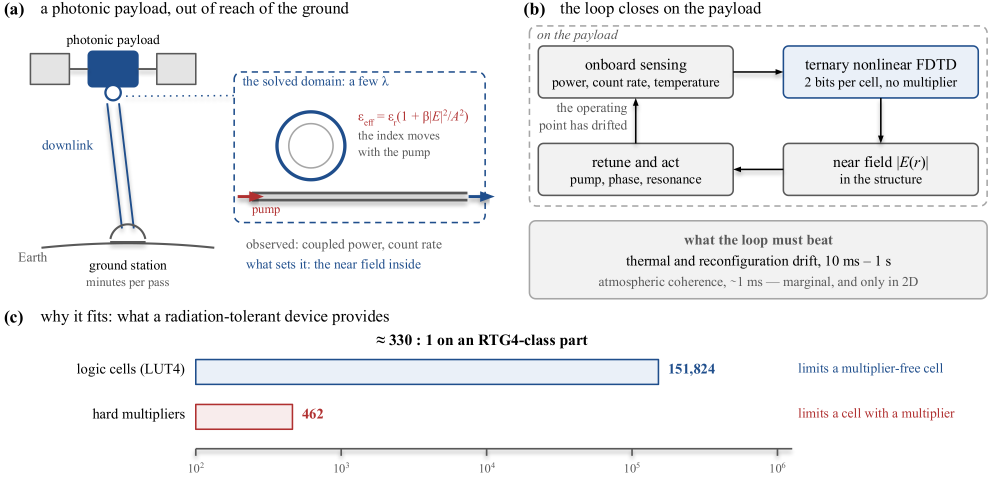}
\caption{\label{fig:onboard}%
  A nonlinear optical solver small enough to fly. (a)~A photonic payload whose operating point is set by a field it cannot measure:~the instrument observes a coupled power or a count rate, while the quantity that determines the correct setting is the near field inside a structure a few wavelengths across, here a Kerr ring whose index moves with the pump. The ground is reachable for minutes per pass.
  (b)~With the solver on the payload, sensing, solution and retuning form one loop, re-entered as the operating point drifts. The timescales it must beat span thermal and reconfiguration drift comfortably and atmospheric coherence only marginally. (c)~Why it fits. A radiation-tolerant device provides its two arithmetic resources in sharply unequal measure---$1.5\times10^{5}$ four-input lookup tables against $462$ hard multipliers~\cite{RTG4}, a ratio of approximately $330$ to $1$. A cell requiring a multiplier is capped by the scarce resource however much logic remains; the multiplier-free cell of Sec.~\ref{sec:3d} is capped only by the abundant one.}
\end{figure*}

One particular motivation for a multiplier-free solver is not that multipliers are expensive on a desktop computer. It is that there are optical systems in which the field must be known, a decision must follow from it and the computer that could supply it is not reachable on the timescale of the decision. Figure~\ref{fig:onboard} sketches the situation. A photonic payload~\cite{Butt24}---a free-space optical terminal~\cite{Kaushal17}, a reconfigurable metasurface~\cite{Shaltout19} or an integrated nonlinear source~\cite{Caspani17}---must hold an operating point that depends on a field distribution inside a wavelength-scale structure~\cite{Mol18}. That distribution is not directly
measurable:~the instrument observes an integrated quantity, a coupled power or a count rate, while the quantity that determines the correct setting is the near field in the structure itself. The payload does not need a map of its own optics;~it needs the answer to a forward problem, repeatedly, as the operating point drifts.

Routing that forward problem to the ground turns a computational problem into one of communication and control latency. A spacecraft can, of course, communicate during a pass, so the objection is not availability but loop delay:~propagation, scheduling and ground processing together set a floor on the round trip that is long compared with the drift it would have to correct, and the link is intermittent between passes~\cite{Kodheli21}. An onboard solver removes that dependency, provided its throughput meets the required response time---which is a question of what the solver costs in hardware that can survive the environment.

The two-bit representation is particularly advantageous here, because the resource a cell needs is the one such devices have in abundance. As established in Sec.~\ref{sec:3d}, the arithmetic of a cell reduces to a selection among a fixed set of coefficient multiples and a table lookup, so a cell consumes logic and memory but no multiplier. That distinction matters because hardened devices are provisioned very unevenly in the two. Radiation-tolerant FPGAs such as Microchip RTG4, a flash-based family qualified for the space radiation environment, offers some $1.5\times10^{5}$ four-input lookup tables against $462$ hard multipliers~\cite{RTG4}, a ratio of about $330$ to $1$. This disparity makes the dedicated arithmetic resources a potential bottleneck in highly parallel designs, whereas the more numerous logic elements offer greater scope for implementing multiplier-free arithmetic.

A conventional implementation need not dedicate a multiplier to each cell, since multipliers can be shared by time multiplexing. However, sharing trades throughput for area, so at a fixed update rate the comparison is between equal numbers of arithmetic units. Additionally, a design requiring one multiplier per unit would be confined to a few hundred units on such a device however much logic remained unused. A multiplier-free unit removes that particular ceiling, leaving replication governed by logic, memory and routing. The design reported in Sec.~\ref{sec:impl} contains no DSP used as a multiplier, confirmed both by inspection of the synthesised connectivity list and by the placed design, in which every multiplier-bearing primitive of the architecture is reported unused (Table~\ref{tab:util}).

Whether that suffices is an arithmetic question worth posing precisely and the figures below are an illustrative budget rather than a performance claim. With $\Delta x=\lambda/P$ and
$\Delta t=\CFL\,\Delta x/c$, one optical period occupies $P/\CFL$ time steps. Writing $N_{c}$ for the number of cells, $n_{p}$ for the number of periods simulated and $N_{p}$ for the number of cells updated per clock at frequency $f_{\mathrm{clk}}$,
\begin{equation}
  T_{\mathrm{solve}}
  = \frac{N_{c}\,n_{p}P}{\CFL\,f_{\mathrm{clk}}N_{p}}\,,
  \qquad
  T_{\mathrm{solve}} \lesssim \tau_{c}\,,
  \label{eq:budget}
\end{equation}
where $\tau_{c}$ is the time over which the tracked quantity remains valid. A two-dimensional section four wavelengths on a side at $P=20$ gives $N_{c}=6.4\times10^{3}$; with $n_{p}=40$, $\CFL=1/32$, $f_{\mathrm{clk}}=250$~MHz and $N_{p}=256$, Eq.~\eqref{eq:budget} gives $T_{\mathrm{solve}}\approx2.6$~ms. Triplicating the datapath for single-event tolerance gives $8$~ms. The Courant number assumed here is eight times coarser than the operating point of Sec.~\ref{sec:optimum}, so the budget buys speed at the cost of oversampling ratio and the noise floor at $\CFL=1/32$ has not been characterised (this is one reason the figures are offered as a budget and not as a performance claim). Thermal and reconfiguration drift, of order $10^{-2}$ to $1$~s, are met with margin. Atmospheric coherence times relevant to adaptive optics are typically of the order of milliseconds, but vary substantially with observing conditions~\cite{Gre77}, and I do not claim that this budget meets them. In fact, the comparison is offered to locate the scheme among optical timescales, not to assert that a turbulence-tracking loop is within reach.

The nonlinear capability is central rather than incidental to these applications. A Kerr medium is one whose index depends on the intensity of the incident light, so the operating point of any $\chi^{(3)}$ device shifts when the pump power changes, and the forward problem must be re-solved rather than looked up. Third harmonic generation, demonstrated in Sec.~\ref{sec:results}, and four-wave mixing, the workhorse of integrated nonlinear photonics (see Ref.~\cite{Mak13} and references therein), are governed by the same susceptibility and a solver that captures that response can track the optical intensity-dependent index and the resonance shifts it produces.

For instance, predicting spontaneous photon pair generation itself requires a quantum or stochastic treatment that a deterministic classical field solver does not supply~\cite{Sol17}. However, a classical solver computes the dispersion, phase matching and detuning on which such a treatment depends~\cite{Ger07}. That the runtime arithmetic is integer throughout---accumulators, error registers, coefficient tables and source table alike, with floating-point confined to table generation before synthesis---is a further advantage in a qualification context:~the datapath is bit-exact and reproducible across devices and across time, so a result obtained on the bench is the result obtained in orbit, which floating-point pipelines do not guarantee. This bears on arithmetic reproducibility only and displaces neither radiation-fault mitigation nor in-orbit validation.

Nevertheless, two things separate the present implementation from the situation sketched in Fig.~\ref{fig:onboard}. The solver demonstrated here is one-dimensional, so the nine-way selection of Sec.~\ref{sec:3d} has been derived but not placed and routed. Additionally, its datapath is deliberately serial, one arithmetic unit shared across the grid, so the replication factor $N_{p}$ of Eq.~\eqref{eq:budget} is an architectural projection.

Neither, however, touches the property the application depends on. What removes the multiplier is the width of the operand rather than the dimensionality of the grid:~going to 3D widens the selection from five ways to nine, Eq.~\eqref{eq:drange} against Eq.~\eqref{eq:fiveway}, at a cost of one addition per cell and no multiplier at all. The serial datapath is likewise a choice and not a constraint, made to fit a part of twenty thousand lookup tables, and the resource asymmetry set out above is what decides how far it replicates on a larger one. The result measured here---that an
inhomogeneous, nonlinear optical FDTD update can be carried out with no multiplier of any kind, on a device chosen for being small---is the premise on which an onboard solver would rest, and it is a measurement rather than an expectation.

\section{Conclusions}
A finite-difference time-domain solver whose field variables are restricted to the ternary alphabet $\{-1,0,+1\}$ reproduces nonlinear optical physics quantitatively and does so on hardware chosen for being small. Starting from the general ternary-valued algorithm, in which each Yee cell holds an integer state that accumulates the discrete curl of neighbouring symbols and is never drained, I have extended the scheme to inhomogeneous and Kerr media and to first-order
Mur absorbing boundaries, in each case without a multiplication. Dielectric contrast becomes a multiplexer, because the difference of ternary symbols takes five values in one dimension and nine in three. In turn, the instantaneous Kerr nonlinearity becomes a coefficient look-up addressed by the local field magnitude and the boundary coefficients become signed-digit shift--add chains. Over $8\times10^{5}$ consecutive time steps captured from the device, the third-harmonic conversion efficiency agrees with a double-precision reference to within $1.4\%$ and the even harmonics that centrosymmetry forbids lie $59$--$65$~dB below the pump against $89$--$94$~dB for the reference. The complete solver contains no multiplier anywhere in its datapath.

Three particular conclusions follow.

First, a two-bit field variable is sufficient for quantitative nonlinear optics. The third-harmonic conversion efficiency is reproduced to $1.4\%$ and the cascade to $5f_{0}$ appears with it.

Second, the restriction to a small alphabet is in part an advantage. A dielectric contrast and an instantaneous Kerr nonlinearity would each require a multiplier in a conventional solver, but here the multiplicand is a difference of ternary symbols taking only five values in one dimension and nine in three, so the product collapses to a multiplexer and the field-dependent coefficient to a table. The alphabet that limits the per-cell information is the same property that eliminates the arithmetic.

Third, the error structure of the nonlinear problem differs
qualitatively from the linear one. Because the state is never drained,
quantisation noise accumulates in proportion to the number of steps and the Courant number trades discretisation accuracy against noise. The existence of an optimum, and the fact that it is not at the smallest accessible $\CFL$, is not anticipated by the linear analysis.

A methodological point follows from the second of these. Because the hardware and the reference evaluate the same integer recurrence, a bit-exact comparison between them cannot detect an error they hold in common:~the floored feedback shift of Sec.~\ref{sec:forbid} passed all four verification layers of Sec.~\ref{sec:impl} and was exposed only by a symmetry of the governing physics. A reference-free invariant is therefore not a redundant check alongside a golden model but a check of a different kind and the forbidden even harmonics supply one at no cost.

The scheme's natural application is to problems where many cells must be updated with modest per-cell precision and where inter-cell bandwidth, not arithmetic, is the constraint~\cite{Mansoori, Pfeifle2015}. The results presented in this paper suggest that the relevant precision budget is set not by the field amplitude but by the oversampling ratio required to keep the shaped noise below the weakest physical signal of interest---here a third harmonic $15$~dB below the pump, resolved with $40$~dB to spare.

%\begin{acknowledgments}
%The design was built entirely with the open-source FPGA toolchain (yosys, nextpnr, Apicula, openFPGALoader).
%\end{acknowledgments}

\section*{Data availability}
The source code is available in the repository cited in
Ref.~\cite{Maksymov2026TernaryFDTD_NLO}.

\section*{Conflict of interest}
The author states no conflict of interest.

\bibliographystyle{apsrev4-2}
\bibliography{biblio}

\end{document}